\documentclass[amsmath,amssymb,superscriptaddress,nobalancelastpage,aps,twocolumn]{revtex4-1}

\usepackage{graphicx}
\usepackage{varioref}
\usepackage{xr-hyper}
\usepackage{xcolor}
\usepackage{nicefrac}
\usepackage{xfrac}
\usepackage{hyperref}
\hypersetup{colorlinks,linkcolor=blue,urlcolor=blue,citecolor=blue}
\usepackage[normalem]{ulem}
\usepackage{lineno}
\usepackage{amsmath}
\usepackage{amssymb}
\usepackage{titlesec}

\emergencystretch=\maxdimen
\begin{document}

\title{On the superconducting gap structure of the miassite Rh$_{17}$S$_{15}$: Nodal or nodeless?}

\author{J. Y. Nie}
\affiliation{State Key Laboratory of Surface Physics, Department of Physics, Fudan University, Shanghai 200438, China}
\affiliation{Shanghai Research Center for Quantum Sciences, Shanghai, 201315, China}

\author{C. C. Zhao}
\affiliation{State Key Laboratory of Surface Physics, Department of Physics, Fudan University, Shanghai 200438, China}

\author{C. Q. Xu}
\affiliation{School of Physical Science and Techonoloy, Ningbo University, Ningbo 315211, China}

\author{B. Li}
\affiliation{School of Science, Nanjing University of Posts and Telecommunications, Nanjing 210023, China}

\author{C. P. Tu}
\affiliation{State Key Laboratory of Surface Physics, Department of Physics, Fudan University, Shanghai 200438, China}

\author{X. Zhang}
\affiliation{State Key Laboratory of Surface Physics, Department of Physics, Fudan University, Shanghai 200438, China}

\author{D. Z. Dai}
\affiliation{State Key Laboratory of Surface Physics, Department of Physics, Fudan University, Shanghai 200438, China}

\author{H. R. Wang}
\affiliation{State Key Laboratory of Surface Physics, Department of Physics, Fudan University, Shanghai 200438, China}

\author{S. Xu}
\affiliation{Department of Applied Physics, Zhejiang University of Technology, Hangzhou 310023, China}

\author{Wenhe Jiao}
\affiliation{Department of Applied Physics, Zhejiang University of Technology, Hangzhou 310023, China}

\author{B. M. Wang}
\affiliation{School of Physical Science and Techonoloy, Ningbo University, Ningbo 315211, China}

\author{Zhu'an Xu}
\affiliation{School of Physics, Zhejiang University, Hangzhou 310058, China}
\affiliation{State Key Laboratory of Silicon and Advanced Semiconductor Materials, Zhejiang University, Hangzhou 310027, China}
\affiliation{Hefei National Laboratory, Hefei 230088, China}

\author{Xiaofeng Xu}
\email{xuxiaofeng@zjut.edu.cn}
\affiliation{Department of Applied Physics, Zhejiang University of Technology, Hangzhou 310023, China}

\author{S. Y. Li}
\email{shiyan\_li@fudan.edu.cn}
\affiliation{State Key Laboratory of Surface Physics, Department of Physics, Fudan University, Shanghai 200438, China}
\affiliation{Shanghai Research Center for Quantum Sciences, Shanghai, 201315, China}
\affiliation{Collaborative Innovation Center of Advanced Microstructures, Nanjing 210093, China}

\date{\today}

\begin{abstract}
Recent penetration depth measurement claimed the observation of unconventional superconductivity in the miassite Rh$_{17}$S$_{15}$ single crystals, evidenced by the linear-in-temperature penetration depth at low temperatures, thereby arguing for the presence of the lines of node in its superconducting gap structure. Here we measure the thermal conductivity of Rh$_{17}$S$_{15}$ single crystals down to 110 mK and up to a field of 8 T ($\simeq 0.4H{\rm_{c2}}$). In marked contrast to the penetration depth measurement, we observe a negligible residual linear term $\kappa_0/T$ in zero field, in line with the nodeless gap structure. The field dependence of $\kappa_0(H)/T$ shows a profile that is more consistent with either a highly anisotropic gap structure or multiple nodeless gaps with significantly different magnitudes. Moreover, first-principles calculations give two electronic bands with complex shape of Fermi surfaces. These results suggest multigap nodeless superconductivity in this multiband Rh$_{17}$S$_{15}$ superconductor.
\end{abstract}

\maketitle
\section{INTRODUCTION}
The observation of heavy-fermion state in correlated $d$-electron systems has attracted extensive research interest in the past three decades~\cite{Kondo1997Li,LiV2O4-Krimmel,LiV2O4-Coleman,NatMater-LiV2O4}. Unlike the $f$-electron counterparts, $d$-electron based heavy fermions are extremely rare and have as yet been reported only in several compounds~\cite{Kondo1997Li,Pinettes1994A}. One of the most notable examples in this class is the heavy-fermion state in the spinel-type transition-metal oxide LiV$_2$O$_4$, the origin of which remains under debate and the magnetic frustration is generally believed to play a prominent role in its formation~\cite{LiV2O4-Coleman,LiV2O4-Urano}. Likewise, the recent discovery of enhanced electron mass in the binary alloy Rh$_{17}$S$_{15}$ has inspired immediate interest since neither of its constituent elements is magnetic and geometric frustration is thought to be weak in its structure~\cite{Matthias1954superconducting,Naren2008strongly,Marzari,Uhlarz,Koyama2010electron,Settai2009Superconducting,Naren2011superconductivity,Naren2011normal,Fukui,Daou,Murakami2020Elastic}.

The rhodium surfide Rh$_{17}$S$_{15}$, also known as miassite, is a mineral existing in nature in the placers of Miass river and thus got its name after that~\cite{Kim2023Unconventional}. The physical properties of this material are exceptional in many senses. First, the heat capacity measurement revealed a significantly enhanced electronic contribution, $\gamma$ = 105 mJ/mol K$^2$, which is a factor of 5 larger than that from the band structure calculations, indicating strong electron correlations in this system~\cite{Naren2008strongly,Marzari,Koyama2010electron}. Second, resistivity of Rh$_{17}$S$_{15}$ displays a broad hump around $\sim$60 K, at which temperature the Hall coefficient changes sign~\cite{Naren2008strongly}. With decreasing temperature, it undergoes a superconducting transition below $T_c$ $\sim$ 5.0-5.5 K, possibly depending on the stoichiometry of the specimens~\cite{Naren2008strongly,Settai2009Superconducting,Naren2011normal,Uhlarz}. Third, $A$ coefficient in its low-$T$ resistivity ($\rho$ $\sim$ $AT^2$) and the Pauli spin susceptibility ($\chi_P$) are both found to be much enhanced compared with those in conventional transition-metal alloys, leading to the Wilson's ratio of 2 that indicates strong correlations in this system~\cite{Naren2008strongly,Wilson_ratio_PRB}. Last, its superconductivity is characterized by a large heat capacity jump at $T_c$ ($\Delta C$/$\gamma T_c$ = 2) and a large upper critical field $H_{c2}$ that exceeds the Pauli limit by a factor of 2~\cite{Naren2008strongly,Naren2011normal}. All these distinctive features differ from a conventional BCS superconductor and therefore place Rh$_{17}$S$_{15}$ in the category of possible unconventional superconductors.

Although much effort has been devoted to understanding the origin of these novel features, the pairing mechanism, especially its superconducting gap symmetry, remains elusive. Previously, the $^{103}$Rh nuclear magnetic resonance (NMR) showed a reduction of the spin part of the Knight shift and an exponential decrease of 1/$T_1$ below $T_c$, suggestive of the spin singlet pairing with an isotropic gap~\cite{Koyama2010electron}. Later on, the specific heat measurement on single crystals by Naren $\textit{et al.}$ pointed out that the specific heat data cannot be fitted by a single $s$-wave gap~\cite{Naren2011normal}. Recently, the $s$-wave gap symmetry in Rh$_{17}$S$_{15}$ was challenged by the London penetration depth measurement, which clearly demonstrated the $T$-linear $\Delta \lambda$ at low temperatures, coincident with the line nodes in the superconducting gap~\cite{Kim2023Unconventional}. More experiments are highly desired to clarify the superconducting gap symmetry of Rh$_{17}$S$_{15}$.

In this Article, we present a detailed low-temperature heat transport study on Rh$_{17}$S$_{15}$ single crystals down to 110 mK and up to a field of 8 T ($\simeq0.4H{\rm_{c2}}$). In zero field, $\kappa_0/T$ extrapolated to zero temperature is essentially zero, which is evidence for nodeless superconducting gaps. The field dependence of $\kappa_0/T$ exhibits a rapid increase at low field, indicating a very anisotropic gap or multiple $s$-wave gaps with significantly different magnitudes. Our results demonstrate nodeless superconductivity in miassite Rh$_{17}$S$_{15}$ with complex Fermi surface topology as shown by band structure calculations, in marked contrast with the nodal superconductivity claimed by the penetration depth measurements.

\section{METHODS}
Single crystals of Rh$_{17}$S$_{15}$ were grown by using a high-temperature flux growth technique~\cite{Kim2023Unconventional}. The x-ray diffraction (XRD) measurement was performed by an x-ray diffractometer (D8 Discover, Bruker). The DC magnetization measurement was performed down to 1.8 K using a magnetic property measurement system (MPMS, Quantum Design). The specific heat was measured in the physical property measurement system (PPMS, Quantum Design) via the long relaxation method. The sample for transport measurements was cut into a rectangular shape with dimensions of 1.69 $\times$ 0.55$ \times$ 0.15 mm$^3$. Four silver wires were attached to the sample with silver paint, which were used for both the resistivity and thermal conductivity measurements. The electrical and heat currents were applied in the (111) plane. The contacts were metallic with a resistance of $\sim$100 m$\Omega$ at 2 K. The in-plane resistivity was measured in the PPMS. The in-plane thermal conductivity was measured in a dilution refrigerator by using a standard four-wire steady-state method with two RuO$_2$ chip thermometers, calibrated $in$ $situ$ against a reference RuO$_2$ thermometer. Magnetic field was applied perpendicular to the (111) plane. To ensure a homogeneous field distribution in the sample, all fields for resistivity and thermal conductivity measurements were applied at a temperature above $T${$\rm_c$}.

The first-principles density functional theory (DFT) calculations were performed by the full-potential linearized augmented plane wave (FP-LAPW) method implemented in WIEN2k package~\cite{wien}. The Perdew-Burke-Ernzerhof (PBE) functional~\cite{pbe} was used for exchange-correlation. The muffin tin radius was set to 2.0 a.u.\ for both Rh and S atoms. $R \cdot K_{max}$ = 7.0  was used for basis set cutoff, where $R$ is the smallest atomic sphere radius and $K_{max}$ is the largest $K$-vector. A $k$-point mesh of 10 $\times$ 10 $\times$ 10 was used to sample the reducible Brillouin Zone (BZ) used for the self-consistent calculation. The Fermi surfaces (FS) were generated using a denser $k$-point mesh of 28 $\times$ 28 $\times$ 28. FS sheets were visualized using the Fermisurfer software~\cite{fermi}.

\section{RESULTS AND DISCUSSION}

\begin{figure}
	\includegraphics[width=8.6cm]{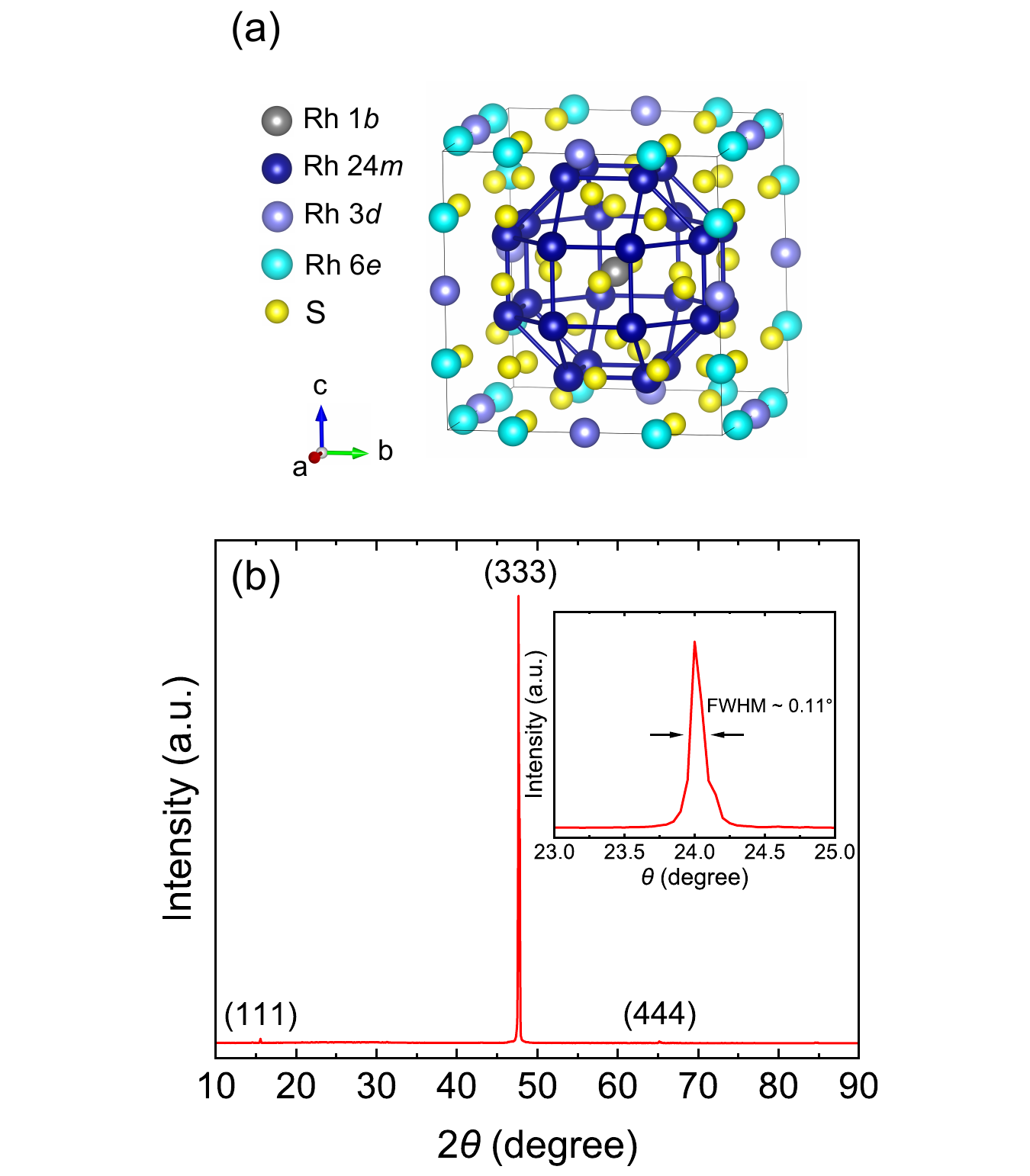}
	\caption{(a) Crystal structure of Rh$_{17}$S$_{15}$. The unit cell has four inequivalent Rh sites (1$\it{b}$, 24$\it{m}$, 3$\it{d}$, and 6$\it{e}$). (b) X-ray diffraction pattern from the largest surface of the Rh$_{17}$S$_{15}$ single crystal, which is identified to be the (111) plane. The inset shows the rocking curve of the (333) reflection with a full-width-at-half-maxima (FWHM) of $\sim$0.11$^{\circ}$. }
	\label{fig.1}
\end{figure}

Rh$_{17}$S$_{15}$ crystallizes in the cubic space group $\it{Pm}$3$\it{m}$. As illustrated in Fig. 1(a), the unit cell of Rh$_{17}$S$_{15}$ consists of two formula units with 64 atoms. Rh atoms occupy four symmetry inequivalent sites (1$\it{b}$, 24$\it{m}$, 3$\it{d}$, and 6$\it{e}$) and S atoms reside at three sites (12$\it{i}$, 12$\it{j}$, and 6$\it{f}$). Among them, Rh 1$\it{b}$ site is located at the body centered position and the Rh 24$\it{m}$ site forms a cage around it. Rh 3$\it{d}$ site resides in the middle of two Rh 6$\it{e}$ sites on the edge of the unit cell.  From x-ray diffraction measurements, as shown in Fig. 1(b), the largest surface of Rh$_{17}$S$_{15}$ single crystal is identified to be the (111) plane. The inset of Fig. 1(b) shows the rocking curve of the (333) reflection with a full-width-at-half-maxima (FWHM) of $\sim$0.11$^{\circ}$, demonstrating high quality of our single crystal. From the obtained interplanar distance, the lattice constant $a = b = c$ is calculated to be 9.916 $\text{\AA}$, which is consistent with previous reports \cite{Uhlarz,Naren2011normal,Daou,Naren2008strongly,Fukui,Settai2009Superconducting}.

\begin{figure}
	\includegraphics[width=8.6cm]{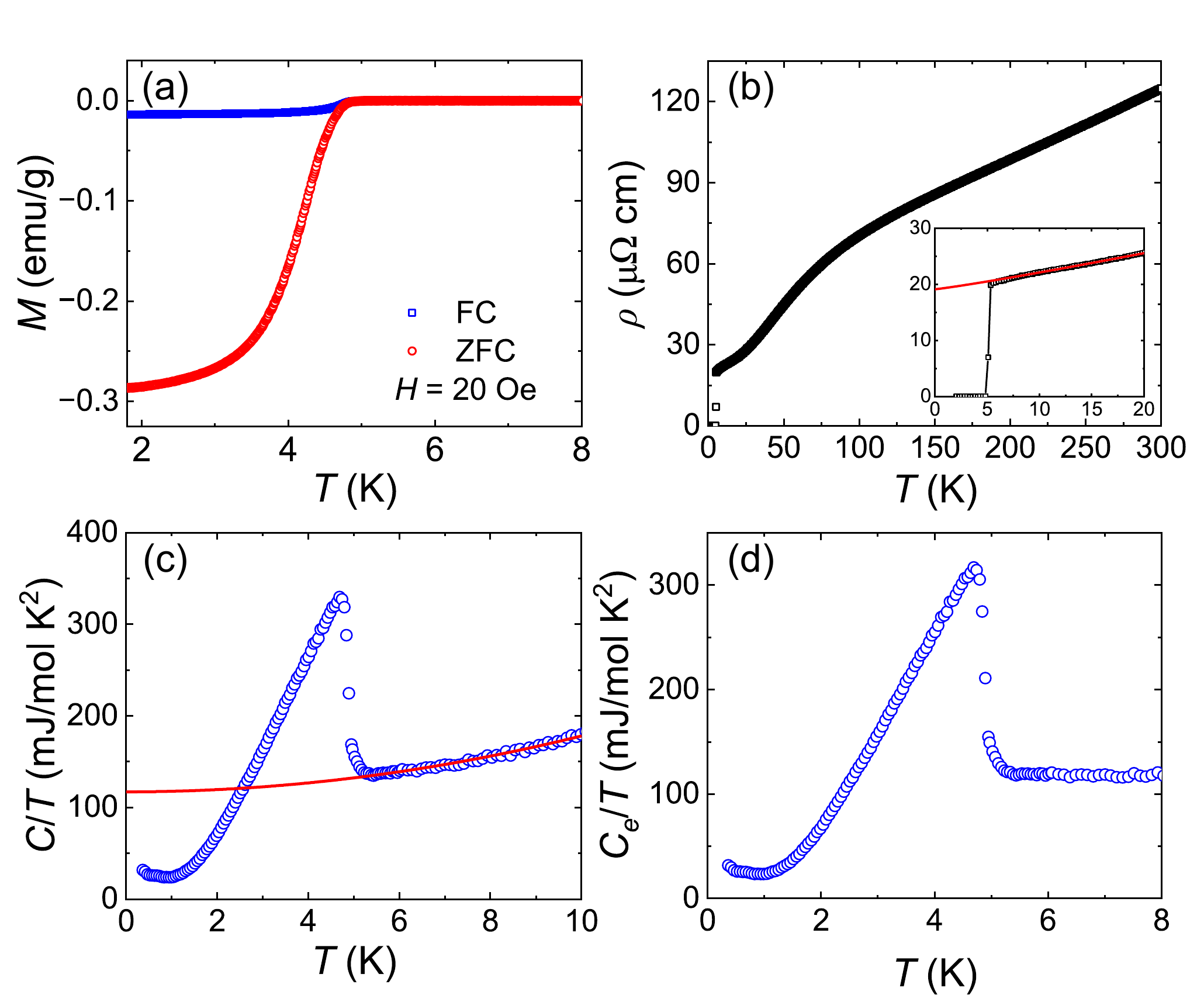}
	\caption{(a) DC magnetization with $\it{H}$ = 20 Oe applied along the $\langle$111$\rangle$ direction of Rh$_{17}$S$_{15}$ single crystal, under both zero-field-cooling and field-cooling modes, respectively. (b) Zero-field resistivity of Rh$_{17}$S$_{15}$. The inset shows the superconducting transition at low temperatures. The solid line shows the fitting to the normal-state resistivity between 8 and 20 K by $\rho (T) = \rho_0 + AT^n$. (c) Temperature dependence of the specific heat divided by temperature $\it{C}$/$\it{T}$ for the Rh$_{17}$S$_{15}$ single crystal at zero field. The red line shows the fitting to $C/T = \gamma + \beta T^{2}$ in the normal state. (d) The electronic specific heat divided by temperature $C_e/T$, after subtracting the phonon term.}
	\label{fig.2}
\end{figure}

Figure 2(a) shows the temperature dependence of the magnetic susceptibility at 20 Oe in both zero-field-cooling and field-cooling modes. The diamagnetic transition occurs at 4.9 K. The temperature dependence of the resistivity $\rho$ at zero field is plotted in Fig. 2(b). The resistivity displays a metallic behavior with a broad knee-like feature around 60 K prior to the superconducting transition, consistent with previous reports~\cite{Naren2008strongly,Settai2009Superconducting,Naren2011normal,Kim2023Unconventional}. The $T\rm_c$ defined by $\rho = 0$ is 5.0 K.
As shown in the inset of Fig. 2(b), the low-temperature resistivity is quite similar to that reported in Ref.~\cite{Settai2009Superconducting}, and the fitting of $\rho (T)$ data between 8 and 20 K to $\rho (T) = \rho_0 + AT^n$ yields the residual resistivity $\rho_0$ = 19.1 $\mu$$\Omega$ cm and $n=1.11$. Figure 2(c) displays the temperature dependence of specific heat divided by temperature $C/T$ in zero field. The specific heat anomaly corresponding to the superconducting transition can be clearly seen at 4.9 K. Above $T\rm_c$, the data from 5.5 K to 10 K can be well fitted by $C/T = \gamma + \beta T^{2}$. The electronic specific-heat coefficient $\gamma$ and the phononic coefficient $\beta$ are determined to be 117.1 mJ mol$^{-1}$ K$^{-2}$ and 0.61 mJ mol$^{-1}$ K$^{-4}$, respectively. The Debye temperature $\Theta\rm_D$ $\approx$ 470 K is estimated by adopting the formula $\Theta\rm_D = (12\pi^4nR/5\beta)^{1/3}$, where universal gas constant $R$ = 8.314 J mol$^{-1}$ K$^{-1}$ and $n$ = 32 is the number of atoms per formula unit. By substracting the phononic contribution, the electronic specific heat is resolved in Fig. 2(d) as $C_{e}/T$ versus $T$. A slight upturn at low temperatures is presumably due to the Schottky anomaly.

\begin{figure}
	\includegraphics[width=8.6cm]{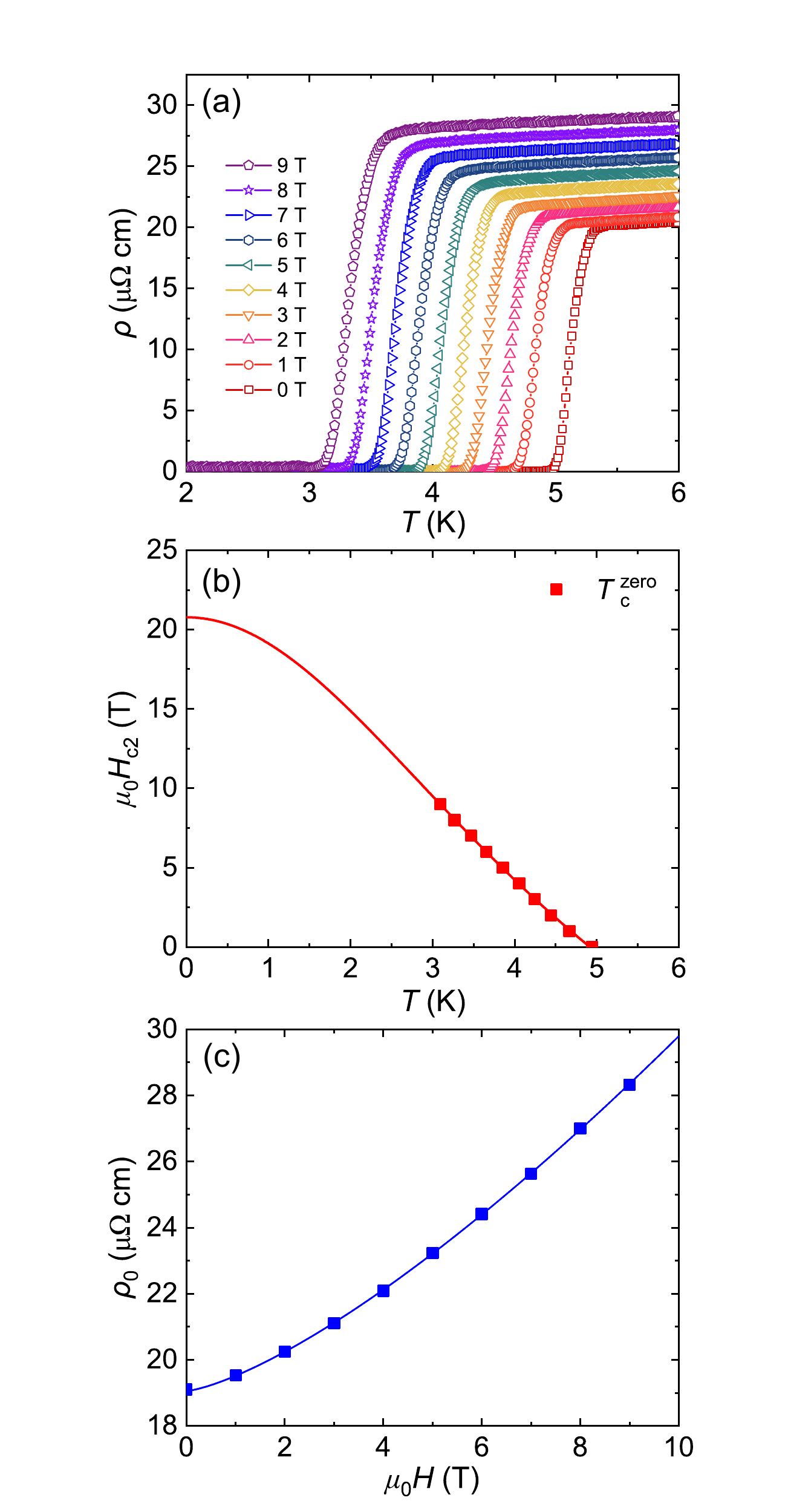}
	\caption{(a) Low-temperature resistivity of Rh$_{17}$S$_{15}$ single crystal under different magnetic fields up to 9 T. (b) Temperature dependence of the upper critical field $\mu_0H{\rm_{c2}}$, extracted from the $T${$\rm_c \rm^{zero}$} values in panel (a). The red line shows the fit to the Ginzburg-Landau formula, from which $\mu_0H{\rm_{c2}}(0) \approx$ 20.8 T is estimated. (c) The field dependence of $\rho_0$. The solid line is a guide to the eye. }
	\label{fig.3}
\end{figure}

Figure 3(a) plots the low-temperature resistivity of Rh$_{17}$S$_{15}$ in magnetic fields up to 9 T, showing that the superconducting transition is gradually suppressed by a magnetic field. The temperature dependence of $H{\rm_{c2}}$, defined by the $T${$\rm_c \rm^{zero}$} values in Fig. 3(a), is plotted in Fig. 3(b). In terms of the Ginzburg-Landau formula, $\mu_0{\it H}_{\rm c2}(\it{T}) = \mu_{\rm 0}{\it H}_{\rm c2}(\rm 0)(1-(\it{T}/{\it T}_{\rm c})^{\rm 2})/(\rm 1+(\it{T}/{\it T}_{\rm c})^{\rm 2})$, the zero-temperature upper critical field $\mu_{\rm 0}{\it H}_{\rm c2}(\rm 0)$ is estimated to be 20.8 T, twice as large as the weak-coupling BCS Pauli limit ($H_P$=$\frac{1.8k_BT_c}{\mu_B})$. Between $\mu_0H$ = 0 and 9 T, the normal state resistivity curves are fitted to obtain the $\rho_0$ for each magnetic field. Figure 3(c) shows the field dependence of $\rho_0$, which exhibits a large magnetoresistance.

\begin{figure}
	\includegraphics[width=8.6cm]{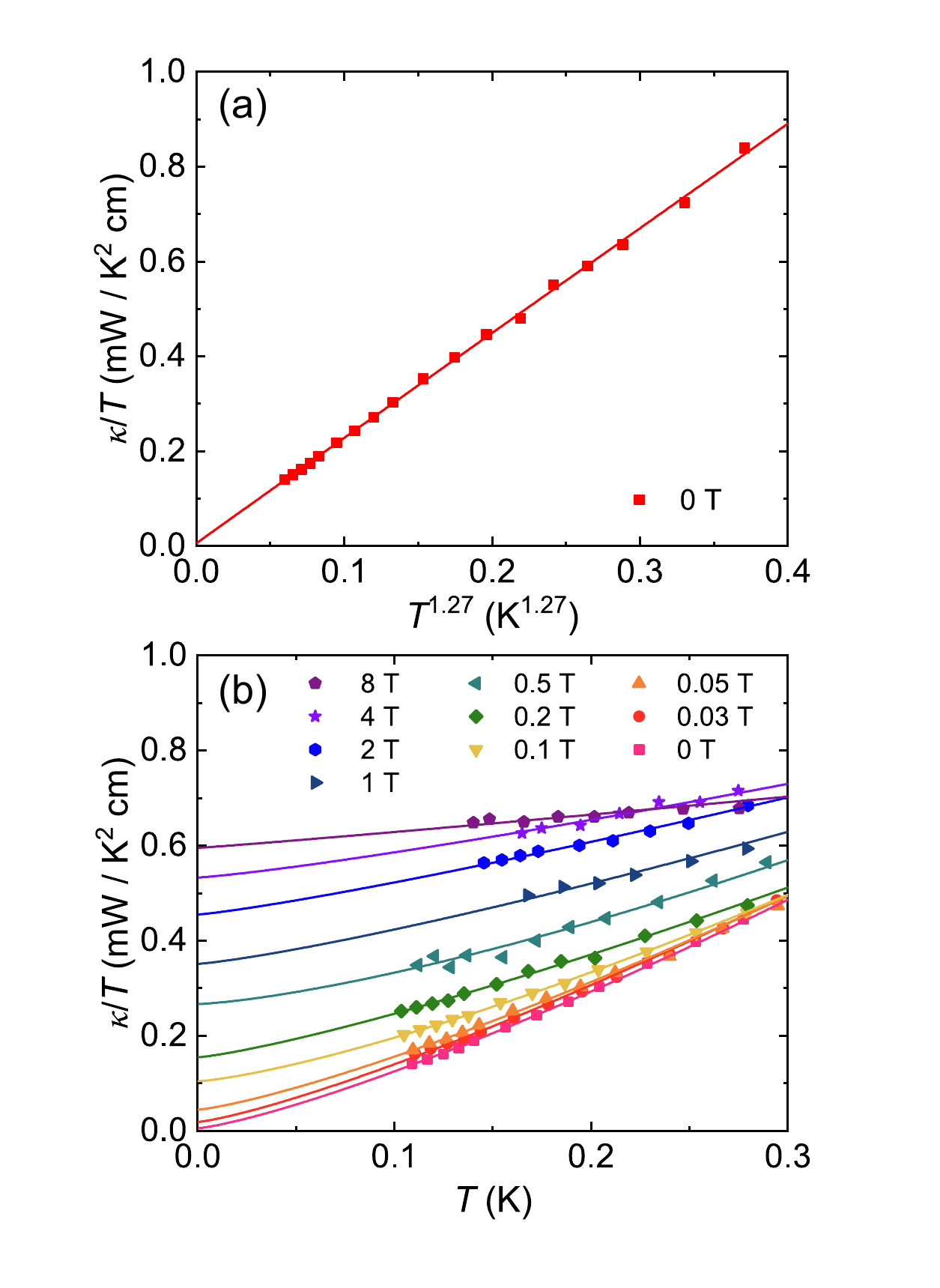}
	\caption{(a) Temperature dependence of the thermal conductivity for the Rh$_{17}$S$_{15}$ single crystal in zero field. The solid line represents a fit of the data to $\kappa/T = a+bT^{\alpha - 1}$, which gives the residual linear term $\kappa_0/T =$ 5 $\pm$ 15 $\mu $W K$^{-2}$ cm$^{-1}$. (b) Low-temperature thermal conductivity of the Rh$_{17}$S$_{15}$ single crystal in magnetic fields up to 8 T. All the curves are fitted to $\kappa/T = a+bT^{\alpha - 1}$, respectively. }
	\label{fig.4}
\end{figure}

Ultra-low-temperature thermal conductivity measurement is a well-established bulk technique that can be used to probe the superconducting gap structure~\cite{Shakeripour2009Heat}. The temperature dependence of thermal conductivity for Rh$_{17}$S$_{15}$ single crystal in zero and magnetic fields is plotted as $\kappa/T$ versus $T$ in Fig. 4. The measured thermal conductivity can be separated into two contributions, $\kappa_e$ and $\kappa_p$, associated with the one from electrons and phonons, respectively. In order to study their specific contributions, the formula $\kappa/T$ = $a$ + $bT^{\alpha-1}$~\cite{Shakeripour2009Heat} is adopted for fitting the data, with the two terms $aT$ and $bT^{\alpha}$ representing contributions from electrons and phonons, respectively. The power $\alpha$ of the second term contributed by phonons is typically between 2 and 3 because of specular reflections of phonons at the sample boundary~\cite{Sutherland2003Thermal,Li2008Low-temperature}.

In zero field, the residual linear term $\kappa_0/T$ of 5 $\pm$ 15 $\mu$W K$^{-2}$ cm$^{-1}$ and the power $\alpha = 2.27$ are obtained by extrapolating the $\kappa/T$ to zero temperature. Compared with our experimental error bar of $\pm$ 5 $\mu$W K$^{-2}$ cm$^{-1}$, the value of $\kappa_0/T$ in zero field is negligible. For an $s$-wave nodeless superconductor, there are no fermionic quasiparticles to conduct heat as $T$ $\rightarrow$ 0, since all electrons are condensed into Cooper pairs~\cite{Shakeripour2009Heat}. Therefore, there is no residual linear term $\kappa_0/T$, as seen in $s$-wave superconductors like Nb, InBi and NbSe$_2$~\cite{Lowell1970Mixed,Willis1976Thermal,Boaknin2003Evidence}. However, for nodal superconductors, a substantial $\kappa_0/T$ in zero field contributed from the nodal quasiparticles has been found. For example, $\kappa_0/T$ of the overdoped ($T_c$ = 15 K) $d$-wave cuprate superconductor Tl$_2$Ba$_2$CuO$_{6+\delta}$(Tl-2201) is 1.41 mW K$^{-2}$ cm$^{-1}$, accounting for $\sim$ 36$\%$ of the normal state $\kappa_{N0}/T$~\cite{Proust2002Heat}. For the superconductor Sr$_2$RuO$_4$ ($T_c$ = 1.5 K), $\kappa_0/T$ = 17 mW K$^{-2}$ cm$^{-1}$ was reported in zero field, more than 9$\%$ of $\kappa_{N0}/T$~\cite{Mackenzie2003The}. Hence, the negligible $\kappa_0/T$ of Rh$_{17}$S$_{15}$ in zero field strongly suggests a nodeless superconducting gap structure.

\begin{figure}
	\includegraphics[width=8.6cm]{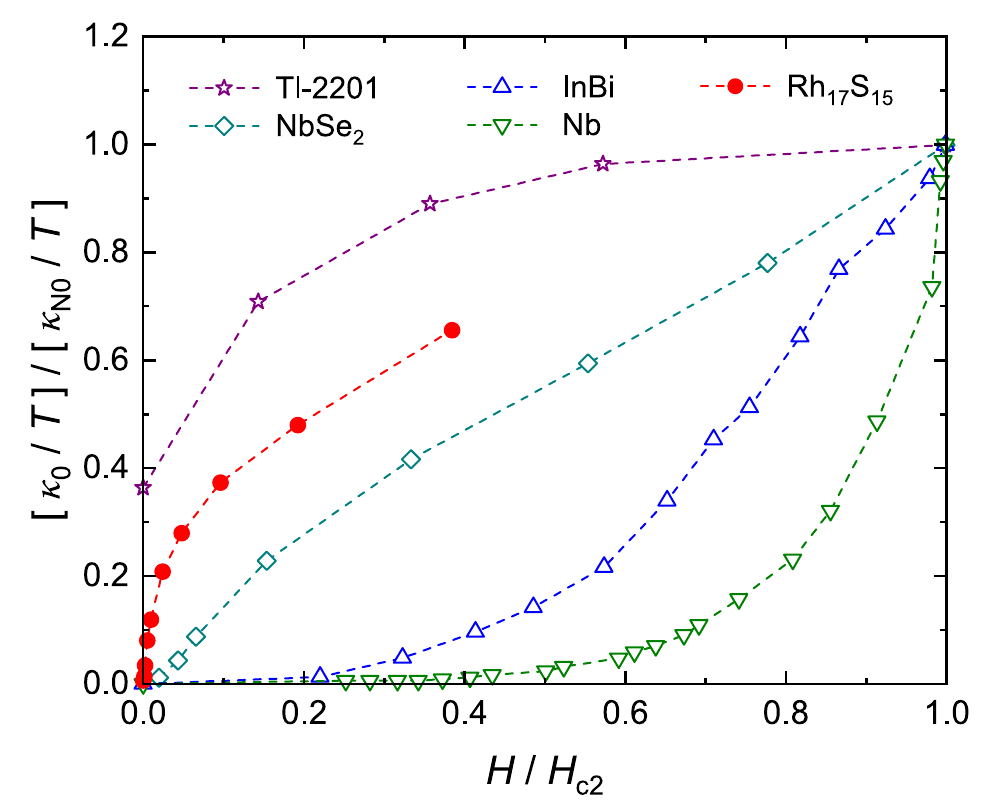}
	\caption{(a) Normalized residual linear term $\kappa_0/T$ of Rh$_{17}$S$_{15}$ as a function of $H/H{\rm_{c2}}$. Data on the clean $s$-wave superconductor Nb~\cite{Lowell1970Mixed}, the dirty $s$-wave superconducting alloy InBi~\cite{Willis1976Thermal}, the multiband $s$-wave superconductor NbSe$_2$~\cite{Boaknin2003Evidence}, and the overdoped $d$-wave cuprate superconductor Tl-2201~\cite{Proust2002Heat} are included for comparison.}
	\label{fig.5}
\end{figure}

\begin{figure*}
	\includegraphics[clip,width=17.2cm]{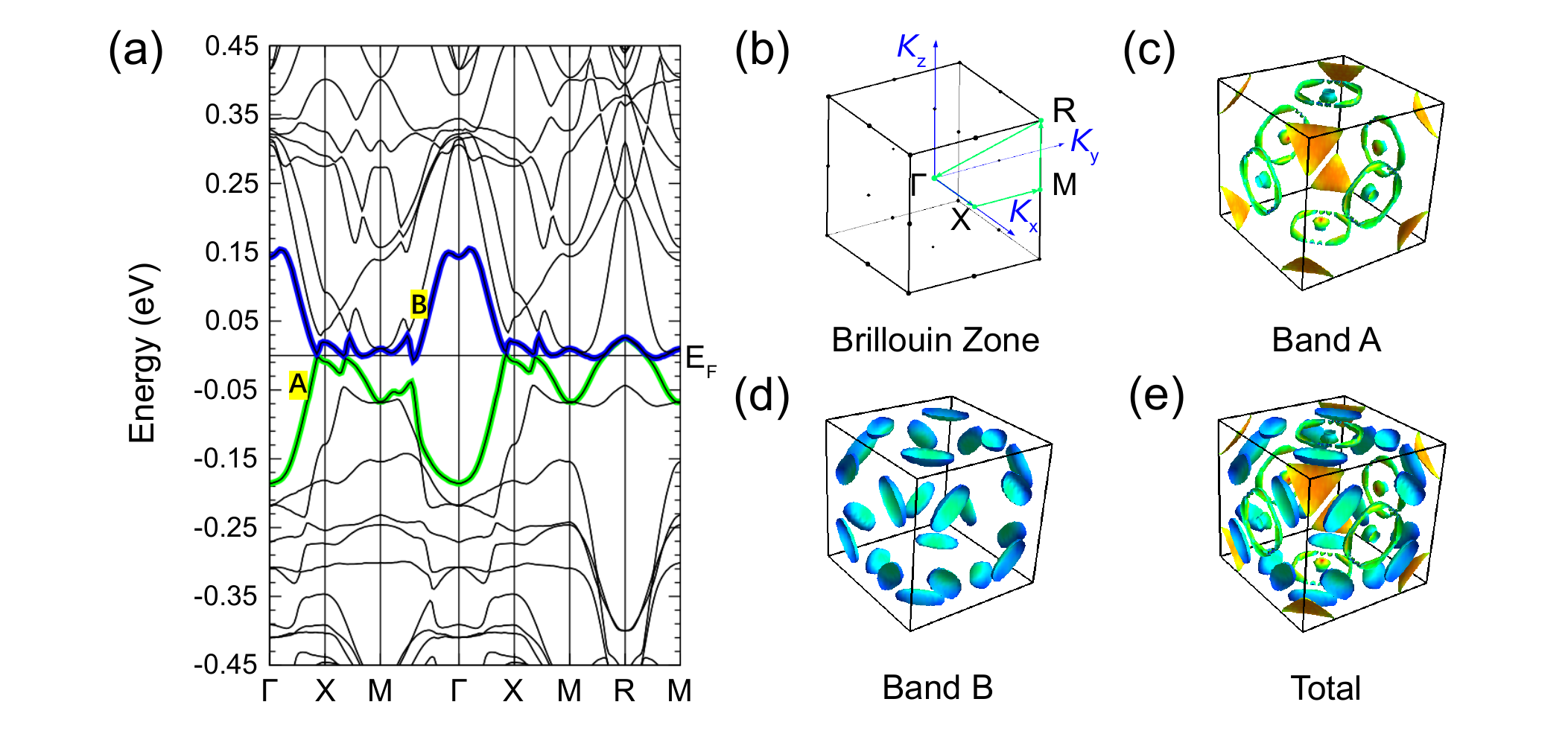}
	\caption{(a) The calculated band structure of Rh$_{17}$S$_{15}$ including the effects of spin-orbit coupling. The Fermi level is set at 0 eV and marked by a horizontal black line. The two bands across the Fermi level $E_F$ are marked as band A and B with different colors in the figure. (b) Brillouin Zone with high-symmetry points and paths. (c)-(d) Partial Fermi surfaces for band A and B. (e) The merged total Fermi surfaces. }
	\label{fig.6}
\end{figure*}
Further insights on the superconducting gap structure can be achieved by examining the profile of field-dependent $\kappa_0(H)/T$~\cite{Shakeripour2009Heat}. As the magnetic field increases, the vortices gradually enter the sample, and the scattering of phonons by electrons becomes more and more dominant, thus the power $\alpha$ of phonon thermal conductivity gradually reduces. We fitted all the curves at different fields with the same formula as did for the zero-field data and thus obtained the $\kappa_0/T$ for each field. Since the magnetoresistance is significant, we use $\kappa{\rm_{N0}}(H)/T \equiv L_0/\rho_0(H)$ to normalize $\kappa_0(H)/T$ for each field. The normalized values as a function of $H/H{\rm_{c2}}$ for Rh$_{17}$S$_{15}$ are plotted in Fig. 5. For comparison, the data for the clean $s$-wave superconductor Nb~\cite{Lowell1970Mixed}, the dirty $s$-wave superconducting alloy InBi~\cite{Willis1976Thermal}, the multiband $s$-wave superconductor NbSe$_2$~\cite{Boaknin2003Evidence}, and the overdoped $d$-wave cuprate superconductor Tl-2201~\cite{Proust2002Heat} are also incorporated. For a clean (like Nb) or dirty (like InBi) type-II $s$-wave superconductor with an isotropic gap, $\kappa_0/T$ grows exponentially with field (above $H{\rm_{c1}}$). This usually gives a negligible $\kappa_0/T$ for fields lower than $H{\rm_{c2}}/4$. For the $d$-wave superconductor Tl-2201, $\kappa_0/T$ increases roughly as $\sqrt H$ at low fields due to the Volovik effect~\cite{Proust2002Heat}. By contrast, for nodeless multiband superconductors, the field dependence of $\kappa_0/T$ depends on the ratio between the large and small gaps. In general, multiband nodeless superconductors have different gap amplitudes, forming approximately two major gaps $\Delta_S$ and $\Delta_L$. Under magnetic fields higher than the characteristic field $H^*\simeq \Delta_s^2$, the superconductivity on the Fermi surface with a smaller gap will firstly be suppressed, and the quasiparticles are then delocalized across the $\Delta_S$, resulting in the enhanced $\kappa_0(H)/T$ in the low-field region, as observed in the multigap superconductor NbSe$_2$~\cite{Kim2021experimental,Boaknin2003Evidence}.

From Fig. 5, the normalized $\kappa_0/T$ of Rh$_{17}$S$_{15}$ starts from a negligible value at zero field, and then increases very rapidly with increasing field. This is a clear indication of a nodeless gap which is either highly anisotropic or of multigap nature, i.e., a small gap on one Fermi surface and a large one on the other~\cite{Boaknin2003Evidence}. Since the Hall resistivity measurements \cite{Naren2008strongly} and band structure calculation \cite{Naren2011normal} clearly demonstrated the multiband character, and the upper critical field shows little anisotropy~\cite{Settai2009Superconducting}, we propose a multigap scenario for Rh$_{17}$S$_{15}$. It is known that the upper critical field is set by $H{\rm_{c2}}(0)\propto(\Delta_0/v_F)^2$~\cite{Bourgeois-Hope2016FeSe}. For the multiband superconductor NbSe$_2$,  the gap on the $\Gamma$ band is approximately one third of the gap on the other two Fermi surfaces and the magnetic field would firstly suppress the superconductivity on the Fermi surface with a smaller gap, resulting in the distinctive shoulder at $H{\rm_{c2}}/9$ in the field dependence of $\kappa_0/T$~\cite{Boaknin2003Evidence}. Similar shoulder is also manifested in MgB$_2$ around $H{\rm_{c2}}/10$~\cite{Sologubenko2002Thermal}.  The even sharper increase in $\kappa_0(H)/T$ in Rh$_{17}$S$_{15}$ may result from an extreme case of multigap structure, in which the gap on one band is much smaller than others.

The calculated electronic band structure and Fermi surfaces of Rh$_{17}$S$_{15}$ are plotted in Fig. 6, including the effects of spin-orbit coupling. We observe two distinct bands crossing the Fermi energy, which are indicated by the different colors in the figure. Notably, there are two Dirac cone-like band touchings that appear between the $\Gamma$-X and X-M high-symmetry paths at the Fermi level. Both electron-like and hole-like bands are present at the Fermi energy, resulting in a multiband hallmark in Rh$_{17}$S$_{15}$. The corresponding Fermi surface in the first Brillouin zone is displayed in Figs. 6(c)-(e). The Fermi surface analysis reveals that band A is hole-like, consisting of 6 rings on the face centers and 8 sheets in the corners. In contrast, band B is electron-like, with 4 ellipsoid-shaped sheets located at each face. This complex Fermi surface topology is consistent with the multiband transport properties evidenced by the Hall resistivity measurements \cite{Naren2008strongly}, which indicate contributions from both electrons and holes in Rh$_{17}$S$_{15}$. It seems quite possible that the gap values for different parts of the Fermi surface differ substantially. Under this circumstance, a small magnetic field would be sufficient to suppress the superconductivity on the weak-gap Fermi surface and result in the rapid increase of $\kappa_0/T$ at low field.

Having established the possible pairing symmetry in Rh$_{17}$S$_{15}$, let us turn to the plausible origin for the disagreement between our measurements and the penetration depth experiment~\cite{Kim2023Unconventional}. The existence of line nodes in the energy gap normally gives the relation of $\Delta \lambda(T)$ $\sim$ $T$ in the clean limit~\cite{hardy1993precision}. However, the negligible $\kappa_0/T$ in zero field clearly disproves this scenario. The rapid increase $\kappa_0/T$ at low field and complex nature of the Fermi surfaces clearly indicate the possible existence of a series of gaps varying in size. It is conceivable that the collective contribution of all the gaps from small to large may give a linear temperature dependence of penetraion depth at low temperature, which mimics the behavior of nodal superconducting gap.

\section{SUMMARY}
In summary, we have investigated the superconducting gap structure of Rh$_{17}$S$_{15}$ single crystals by measuring the low-temperature thermal conductivity down to 110 mK. In contrast to the penetration depth measurement which claims the extended $s$-wave gap with ring-shaped accidental nodes, a negligible residual linear term $\kappa_0/T$ in zero field and the field dependence of $\kappa_0/T$ observed in Rh$_{17}$S$_{15}$, combined with the complex multiple Fermi surfaces from band structure calculations, strongly suggest an extreme case of multigap nodeless superconductivity. Further experimental techniques, such as the angle-resolved photoemission spectroscopy or scanning tunneling microscopy experiments are highly desirable to directly probe the superconducting gap in momentum space in Rh$_{17}$S$_{15}$.

\section*{ACKNOWLEDGMENTS}
This work was supported by the Natural Science Foundation of China (Grant No. 12174064) and the Shanghai Municipal Science and Technology Major Project (Grant No. 2019SHZDZX01). Xiaofeng Xu was supported by the National Natural Science Foundation of China (Grants No. 12274369 and No. 11974061). Zhu'an Xu was supported by the Natural Science Foundation of China (Grant No. 12174334) and the National Key Projects for Research \& Development of China (Grant No. 2019YFA0308602). B. M. Wang and C. Q. Xu were supported by the Zhejiang Provinvial Natural Science Foundation of China (LD24F040001).

\end{document}